\documentclass[12pt]{iopart}

\usepackage{iopams}
\usepackage{epsf,epsfig}
\newcommand{\bq}{\begin{equation}}
\newcommand{\eq}{\end{equation}}
\newcommand{\tab}{\hspace {1.5 em}}

\newcommand{\bqa}{\begin{eqnarray}}
\newcommand{\eqa}{\end{eqnarray}}
\newcommand{\fa}{\forall}
\newcommand{\gdnl}{|GD_N^{(\ell)}\rangle}

\begin{document}
\title{Reducible Correlations in Dicke States}
\author{Preeti Parashar and Swapan Rana}
\address{Physics and Applied Mathematics
Unit, Indian Statistical Institute,\\ 203 B T Road, Kolkata-700
108, India} \ead{parashar@isical.ac.in, swapan\_r@isical.ac.in}
\begin{abstract}
We apply a simple observation to show that the generalized Dicke
states can be determined from their reduced subsystems. In this
framework, it is sufficient to calculate the expression for only
the diagonal elements of the reudced density matrices in terms of
the state coefficients. We prove that the correlation in
generalized Dicke states $\gdnl$ can be reduced to $2\ell$-partite
level. Application to the Quantum Marginal Problem is also
discussed.
\end{abstract}
\pacs{03.67.-a, 03.65.Ud, 03.67.Mn} \submitto{J. Phys. A: Math.
Theor. (\textbf{Fast Track Communication})}

\textbf{\emph{Introduction}}: Entanglement is one of the most
fascinating non-classical  features of quantum theory which has
been harnessed for various  practical applications. Although
bipartite entanglement is well understood,  gaining insight into
multipartite entanglement is still quite a challenge. There are
various  perspectives to study entanglement at the multi-party
level such as  its characterization by means of LOCC, its ability
to reject local realism  and hidden variable theories etc. A
particular interesting point of view is  that of ``parts and
whole". This approach basically deals with the question:
 how  much knowledge about the quantum system can be acquired from that of its
 subsystems? To be precise, it asks whether an unknown state can be determined
 \emph{uniquely} if all its reduced density matrices (RDMs) are specified. In other words,
 this means whether higher order correlations are determined by lower order ones. It
 turns out that the most entangled states are the
 ones which cannot be determined from their RDMs.

The determination of a state from its RDMs implies that the
correlation present in the state is reducible to lower order ones.
In an interesting work \cite{LPW} it was shown that except the GHZ
class ($a|000\rangle+b|111\rangle$), all $3$-qubit pure states are
determined by their $2$-qubit
RDMs. This was further generalized \cite{WL} to the $N$-qubit case
to show that GHZ is the most entangled class of states. In these
works, the knowledge of $(N-1)$-party RDMs was employed to
characterize the $N$-party state. However, in the general scenario
there may exist states which can be determined by less than
$(N-1)$-party RDMs, i.e., a generic correlation can be reduced
beyond the $(N-1)$-partite level. For example, we have recently
shown \cite{PR} that the $N$-qubit $W$ class of states are
determined by just their bipartite RDMs.

Though some partial progress has been made in this direction
\cite{others1}, there is no general technique to know which class
of states can be determined by $K$-partite RDMs for $K<N-1$.
Answering this question will lead to the classification of quantum
states in terms of various kinds of reducible correlations they
can exhibit \cite{LPW, WL, PR}. A natural way to solve the problem
is to determine all the RDMs from the given state and from an
arbitrary state (which is supposed to have the same set of RDMs)
and then compare the corresponding RDMs. But this is practically
very difficult as we need to solve several second degree equations
involving complex numbers.

In this communication we provide some interesting examples of
states which can be determined by their $K$-partite RDMs for $K <
N-1$. In particular, we shall consider the Dicke states which are
genuinely entangled and have been widely studied both from
theoretical and experimental point of views \cite{Dref1}. The
simplest Dicke state is the $W$ state which was studied at the
qubit level recently \cite{PR}. In the present work, as a first
step, we shall extend this result to arbitrary $d$-dimensional
(i.e., $N$-qudit) $W$-state. Next, we shall focuss on the
$N$-qubit Dicke states and study their reducible correlations.
This result is further generalized to $d$-dimensions. Another
interesting application of our technique that would be mentioned
is the Quantum Marginal Problem.

Our proof is based on the simple fact that the RDMs of a
\emph{pure} state can be constructed only from the expressions of
the diagonal elements. This facilitates easy computation of the
RDMs. In addition, if some of the diagonal entries are zero, then
this constraints some diagonals of the arbitrary state to vanish,
thereby reducing the number of unknowns. So first, let us rewrite
some known observations and notational conventions to construct
RDM's, in a slightly different way, for later convenience.

\textbf{Observation }: \emph{To calculate the Reduced Density
Matrix from a generic pure state, it is sufficient to calculate
the expression for diagonal elements in terms of the state
coefficients. All off-diagonal elements will be obtained from
these expressions.}

A density matrix being hermitian, can be identified by its
upper-half elements $a_{ij}~~\fa i\le j$. So we do not need to
calculate the lower-half elements.

Using the `lexicographically-ordered' basis \{$|00...0\rangle$,
$|00...1\rangle$, ... ,$|00...\overline{d-1}\rangle$, ... ,
$|\overline{d-1}~\overline{d-1}...\overline{d-1}\rangle$\} of
${\mathbb{C}^d}^{\otimes N}$, an $N$-qudit pure state $|\psi\rangle_N^d$
(i.e., an $N$-partite pure quantum state where each of the parties has a
$d$-level system) can be expressed as \bq\label{psi1}
|\psi\rangle_N^d=\sum_{i=0}^{d^N-1}c_i|D_N(i)\rangle,~
\sum_{i=0}^{d^N-1}|c_i|^2=1,\eq where $D_N(x)\equiv$
``Representation of the decimal number $x$ in an $N$-bit string in
$d$-base number system". To have a grip on the coefficient
corresponding to a basis vector, we are using $d$-base number
system to represent the basis vector so that the suffix of its
coefficient can be obtained by converting it into decimal number
and vice-versa. [Note that for
$d\geq11$, we need at least two bit (digit) to represent $d-1$ in
decimal number system. But we wish to restrict ourselves to using one bit to
represent one level. So we are using $d$-base number system to represent the bases.
Thats why a `bar' is used over $d-1$ to indicate that it is of $d$-base number system
(and so it consists of one bit)].

Throughout the discussion, we will use $d$-base number system to
represent only the bases  and decimal numbers elsewhere. However,
when there is no ambiguity, we will write $|i\rangle$ instead of
$|D_N(i)\rangle$ -- it should always be understood that the bases
are in $d$-base number system.

Now let us calculate the $M$-partite marginal (RDM) $
\rho^{i_1i_2...i_M}_\psi=Tr(|\psi\rangle_N^d\langle\psi|)$,  where
the trace is taken over the remaining $N-M$ parties. Clearly, it
will be a matrix of order $d^M\times d^M$. So, retaining only the
upper half entries, we can write \bq\label{rhopsi1}
\rho^{i_1i_2...i_M}_\psi~=~\sum_{i=0}^{d^M-1}\sum_{j=i}^{d^M-1}r_{ij}|D_M(i)\rangle\langle
D_M(j)|.\eq Since the RDM is obtained by tracing over $N-M$
parties (the space of these parties has dimension $d^{N-M}$), each
$r_{ij}$ will be a sum of $d^{N-M}$ number of terms each of which
is of the form $c_k\bar{c}_l$. Thus
$r_{ij}=\sum_{p=0}^{d^{N-M}-1}c_{k_p}\bar{c}_{l_p}$. To get the
expression of $r_{ij}$ (i.e., to see explicitly which $c_k$'s and
$c_l$'s will appear in the sum) , let us fix one $i$ and one $j$.
Let $D_M(i)=s_1s_2...s_M$ and $D_M(j)=t_1t_2...t_M$. In an $N$-bit
string, let us now put the $s_j$'s at $i_j$th places respectively.
Then the suffixes $k$'s will be obtained by converting the $N$-bit
$d$-base numbers, obtained by filling all the remaining $N-M$
places of the above string arbitrarily with 0, 1,
2,..,$\overline{d-1}$, into decimal numbers.
\begin{figure}[ht]
\epsfxsize=3in\centerline{\epsfbox{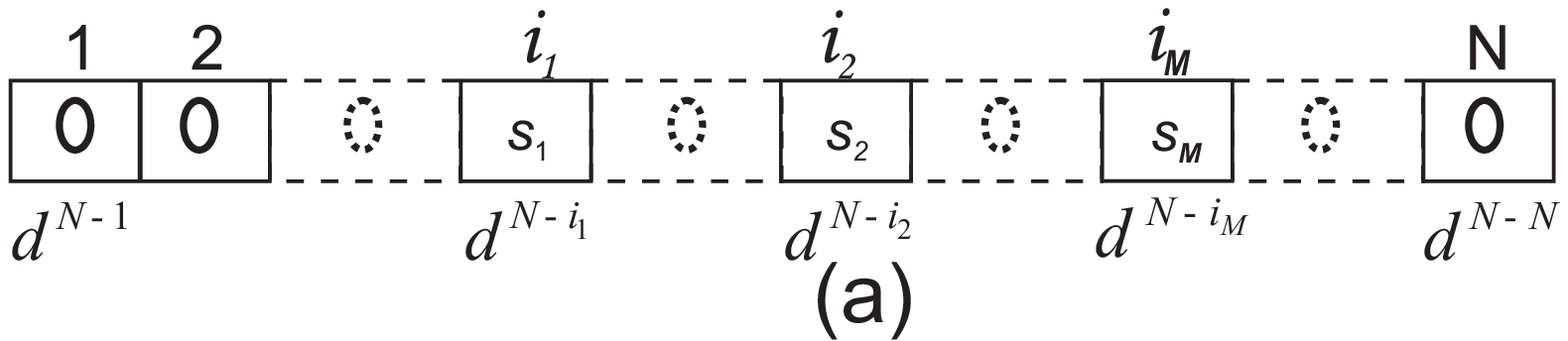}~~\epsfxsize=3in\epsfbox{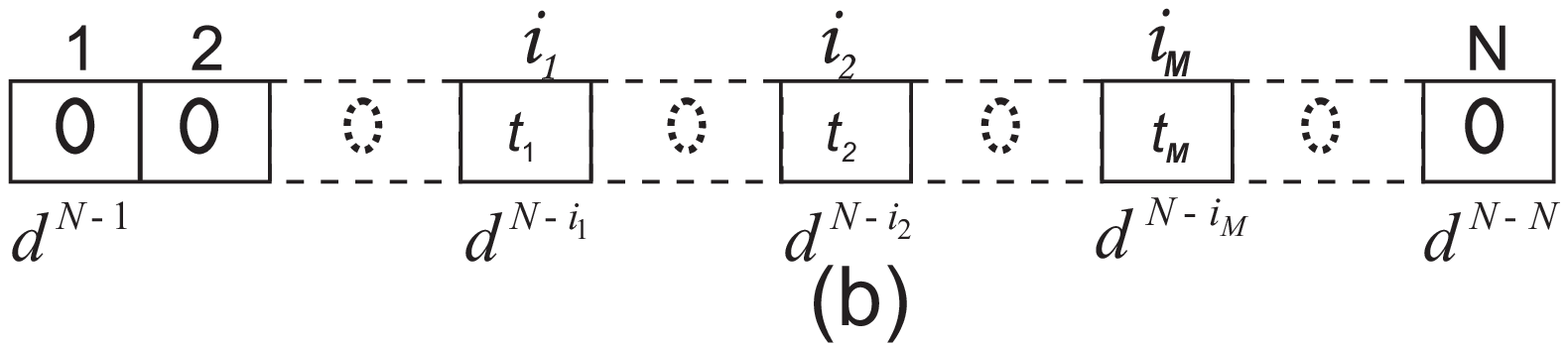}}
\label*{}\caption{Least suffix (or the first term) in $r_{ij}$ is
$c_{k_0}\bar{c}_{l_0}$, \\where (a)
$k_0=\sum_{j=1}^Ms_j.d^{N-i_j}$ and (b)
$l_0=\sum_{j=1}^Mt_j.d^{N-i_j}$}
\end{figure}
For an illustration, let $k_0$ be the decimal number obtained by
converting the $N$-bit $d$-base number having $s_j$ fixed at
$i_j$th place ($\fa j=1(1)M$) and zero at all the remaining $N-M$
places [see fig. 1 for illustration]. Then
$k_0=\sum\limits_{j=1}^Ms_j.d^{N-i_j}$ . Similarly, let
$l_0=\sum\limits_{j=1}^Mt_j.d^{N-i_j}$. Then the first term
(ordering in suffixes appear) of the sum in the expression of
$r_{ij}$ will be $c_{k_0}.\overline{c}_{l_0}$. In a similar way
other $c_{k_p}.\overline{c}_{l_p}~~(p=0(1)(d^{N-M}-1))$ terms can
be calculated. For the last term, which is the term with highest
suffix, we have $k_{d^{N-M}-1} =k_0+(d-1)\sum\limits_{j=1;j\neq
i_1,i_2,..,i_M}^Nd^{N-j}$. Note that
$r_{ii}=\sum\limits_{p=0}^{d^{N-M}}|c_{k_p}|^2$ and
$r_{jj}=\sum\limits_{p=0}^{d^{N-M}}|c_{l_p}|^2$. Since
$r_{ij}=\sum\limits_{p=0}^{d^{N-M}}c_{k_p}.\overline{c}_{l_p}$, it
follows that each off-diagonal element $r_{ij}$ can be obtained by
summing the products of corresponding complex numbers appearing in
the expression of $r_{ii}$ and conjugate of the complex numbers
appearing in the expression of $r_{jj}$. Hence, it is sufficient
to calculate the expression for only the diagonal elements.
\hfill$\Box$

\textbf{\emph{Remark 1}:} If we start with a mixed state
$\left[r_{ij}\right]_{j\ge i=0}^{d^N-1}$ and we wish to calculate
an RDM $\left[R_{ij}\right]_{j\ge i=0}^{d^M-1}$, following the
same procedure, it can be shown that if
$R_{ii}=\sum_{s=0}^{d^M-1}r_{(k_s)(k_s)}$ and
$R_{jj}=\sum_{s=0}^{d^M-1}r_{(l_s)(l_s)}$, then
$R_{ij}=\sum_{s=0}^{d^M-1}r_{(k_s)(l_s)}$. So $R_{ii}=0$ would
imply $r_{(k_s)(j)}=0~~\fa s,j!$  Thus, it is always helpful to
calculate first the expression of the diagonal elements ( and
compare them).

We shall now apply the above observation to arrive at our main
results. As a natural extension of the work on $N$-qubit $W$ state
\cite{PR}, first we shall consider the generalized $d$-dimensional
$W$ state. This would also serve as a good demonstration of the
technique and be useful in understanding the proof for the
generalized Dicke states.

\textbf{\emph{I. $N$-qudit generalized $W$-state}}:

The $N$-qudit generalized $W$-state is defined as
\cite{Sanders}\bq
|W\rangle_N^d=\sum_{i=1}^{d-1}(a_{1i}|i0...00\rangle+...+a_{ni}|00...0i\rangle)
\nonumber \eq However, we will write this state in our notation as
\bq\label{dw1}
|W\rangle_N^d=\sum_{i=0}^{N-1}\sum_{j=1}^{d-1}w_{jd^i}|D_N(jd^i)\rangle
, \sum_{i=0}^{N-1}\sum_{j=1}^{d-1}|w_{jd^i}|^2=1.\eq

\textbf{Theorem 1:} \emph{$N$-qudit generalized $W$-states are
determined by their bipartite marginals.}

We shall prove this by showing that there does not exist any other
$N$-qudit density matrix having the same bipartite marginals
except \bqa\label{dw2} |W\rangle_N^d\langle W|&=&
\sum_{i=0}^{N-1}\sum_{j=1}^{d-1}\sum_{k=j}^{d-1}w_{jd^i}\bar{w}_{kd^i}|D_N(jd^i)\rangle\langle
D_N(kd^i)|\nonumber\\
&+&\sum_{i=0}^{N-1}\sum_{j=1}^{d-1}\sum_{l=i+1}^{N-1}\sum_{m=1}^{d-1}w_{jd^i}\bar{w}_{jd^l}|D_N(jd^i)\rangle\langle
D_N(md^l)|.\eqa [Note that though the above expression looks
rather cumbersome, the matrix form can be easily visualized as
there are non-zero elements only at $(jd^i+1, kd^l+1)$ positions
where $j,k=1(1)(d-1);~i,l=0(1)(N-1)\mbox{ and } jd^i\le kd^l$,
since we are considering only the upper half elements. These
elements are the coefficients of $|D_N(jd^i)\rangle\langle
D_N(kd^l)|$ and are given by $w_{jd^i}\bar{w}_{kd^l}$.]

\textbf{\emph{Proof}:}

1.\tab Each bipartite marginal $\rho^{JK}_W$ will be a matrix of
order $d^2 \times d^2$ where $J \in \left\{1,2,...,N-1\right\}$
and $K \in \left\{2,3,...,N\right\}$. As discussed earlier, to
determine $\rho^{JK}_W$, we need to find the expressions of the
$d^2$ diagonal elements of $\rho^{JK}_W$ i.e., the coefficients of
$|ij\rangle\langle ij|~~\forall i,j=0(1)(d-1)$. [Note that, while
in the bases, $i,j$ should be understood as $d$-base numbers].

  Now each basis state in (\ref{dw1}) has exactly one nonzero entry (rather `bit') with value 1 to
 $d-1$. So there will be no basis term $|ij\rangle\langle ij|$
 of $\rho^{JK}_W$ having both $i,j$ as nonzero numbers. Therefore,
 the coefficient of $|ij\rangle\langle ij|$ in $\rho^{JK}_W$ should
 be zero $\forall i,j=1(1)(d-1)$. Hence we need  to consider only
 the coefficients of $|0i\rangle\langle 0i|$ and $|i0\rangle\langle i0|$~~$\forall
 i=0(1)(d-1)$.

 Again, for $i\neq0$, there is exactly one basis state in (\ref{dw1}) containing `$i$'
 at $J$th place (from left to right) having the coefficient
 $w_{id^{N-J}}$. Therefore, the coefficient of $|i0\rangle\langle
 i0|$ is $|w_{id^{N-J}}|^2$. Similarly, the coefficient of $|0i\rangle\langle 0i|$
 in $\rho^{JK}_W$ is $|w_{id^{N-K}}|^2$~~$\forall
 i=1(1)(d-1)$. From normalization ($Tr(\rho^{JK}_W)=1$), the
 coefficient of $|00\rangle\langle00|$ in $\rho^{JK}$ is obtained
 as $$1-\sum_{i=1}^{d-1}(|w_{id^{N-J}}|^2+|w_{id^{N-K}}|^2).$$

 We know that the non-diagonal terms (the coefficients of
 $|0i\rangle\langle0j|$, $|0i\rangle\langle j0|$ and
 $|i0\rangle\langle j0|$; $i\neq j$) will be determined from the
 above expressions of the diagonal terms. Thus,\\ \bqa \label{gwrdm}\rho^{JK}_W&=&
 (1-\sum_{i=1}^{d-1}(|w_{id^{N-J}}|^2+|w_{id^{N-K}}|^2))|00\rangle\langle00|
 +\sum_{i=1}^{d-1}\sum_{j=i}^{d-1}w_{id^{N-K}}\bar{w}_{jd^{N-K}}|0i\rangle\langle0j|\nonumber\\
 &+&\sum_{i=1}^{d-1}\sum_{j=1}^{d-1}w_{id^{N-K}}\bar{w}_{jd^{N-J}}|0i\rangle\langle j0|
 +\sum_{i=1}^{d-1}\sum_{j=i}^{d-1}w_{id^{N-J}}\bar{w}_{jd^{N-J}}|i0\rangle\langle
 j0| .
 \eqa


2.\tab Now let us suppose that there exists an $N$-qudit density
matrix (possibly mixed, hence the subscript M)\bq
\rho^{12...N}_M~=~\sum_{i=0}^{d^N-1}\sum_{j=i}^{d^N-1}r_{ij}|D_N(i)\rangle\langle
D_N(j)|\eq which has the same bipartite marginals as
$|W\rangle_N^d$. Here $r_{ii}\ge0~\fa i=0(1)(d^N-1)$, since the
diagonal elements of a Positive Semi Definite (PSD) matrix are
non-negative. [If possible let in a PSD matrix A, a diagonal
element $d_i<0$. Then taking $|\psi\rangle
=[0,0,..,0,1,0,...0]^T$, we have
$\langle\psi|A|\psi\rangle=d_i<0$, a contradiction that A is PSD].

We first wish to calculate the diagonal elements of the bipartite
marginal $\rho^{JK}_M$ of $\rho^{12...N}_M$. Each diagonal element
of $\rho^{JK}_M$ is the sum of $d^{N-2}$ number of diagonal
elements $r_{ss}$ of $\rho^{12...N}_M$. Out of the $d^2$ number of
diagonal elements (coefficients of
 $|ij\rangle\langle ij|~~\forall i,j=0(1)(d-1)$) of $\rho^{JK}_M$, let us first calculate
 the coefficient of $|ij\rangle\langle ij|~~\forall
 i,j=1(1)(d-1)$. To see explicitly which $r_{ss}$'s will appear
in the sum, we observe that the suffixes $s$ will vary over the
decimal numbers obtained by converting the $N$-bit $d$-base
numbers having $i$ fixed at $J$th \& $j$ fixed at $K$th places and
arbitrarily 0,1,...,$\overline{d-1}$ at the remaining ($N$-2)
places. Hence the terms $r_{ss}$'s for the suffixes $s=0$ and
$s=k.d^{l-1}, k=1,2,..,(d-1); l=1,2,...,N$ will not appear in the
expression (sum) of coefficient of $|ij\rangle\langle ij|$ in
$\rho^{JK}_M$ for any $J,K$ as for these $s$, $D_N(s)$ can have at
most one non zero entry (but we need at least two).

3a.\tab As can be seen from eqn.(\ref{gwrdm}), there is no term
$|ij\rangle\langle ij|$ for $ij\neq0$ in $\rho^{JK}_W$. Therefore,
the coefficient of $|ij\rangle\langle ij|$ for $ij\neq0$ in
$\rho^{JK}_M$ should vanish. Since these coefficients are sum of
non-negative $r_{ss}$'s, each $r_{ss}$ appearing there should
individually be zero. Therefore, from step (2), the only non-zero
diagonal elements of $\rho^{12...N}_M$ are $r_{ii}$ for $i=0~  \&
~ i=j.d^{k-1}~\forall j=1(1)(d-1),k=1(1)N$.

3b.\tab Next comparing the coefficient of $|0i\rangle\langle 0i|$
from $\rho^{JK}_W$ and $\rho^{JK}_M$, we get
$r_{(id^{N-K})(id^{N-K})}=|w_{id^{N-K}}|^2$ for all $i=1(1)(d-1)$.
Similarly, comparing the coefficient of $|i0\rangle\langle i0|$,
we get $r_{(id^{N-J})(id^{N-J})}=|w_{id^{N-J}}|^2$ for all
$i=1(1)(d-1)$.  Since these results hold for all possible
(parties) $J$ and $K$, we can write them in combined form as
$r_{(jd^i)(jd^i)}=|w_{jd^i}|^2~\forall j=1(1)(d-1),i=0(1)(N-1)$.

3c.\tab Finally, from normalization condition
$\sum\limits_{i=0}^{d^N-1}r_{ii}=1=\sum\limits_{i=0}^{N-1}\sum\limits_{j=1}^{d-1}|w_{jd^i}|^2$,
we get $r_{00}~=~0$. Thus collecting the results of steps 3a \& 3b
it follows that\bq\label{cond1} r_{(jd^i)(jd^i)} = |w_{jd^i}|^2,
\fa j=1(1)(d-1),i=0(1)(N-1)\eq and all other $r_{ii}$ in
$\rho^{12...N}_M$ are zero.

4.\tab Now we will use the fact that if a diagonal element of a
PSD matrix is zero, then all elements in the row and column
containing that element should be zero \cite{PSD2}. Hence from the
result of step 3c it follows that $\rho^{12...N}_M$ has nonzero
elements only at $(jd^i+1, kd^l+1)$ positions where
$j,k=1(1)(d-1);~i,l=0(1)(N-1)\mbox{ and } jd^i\le kd^l$. These
elements are the coefficients of $|D_N(jd^i)\rangle\langle
D_N(kd^l)|$ and are given by $r_{(jd^i)(kd^l)}$. Therefore,
$\rho^{12...N}_M$ has the same form as $|W\rangle_N^d\langle W|$
given in (\ref{dw2}). Moreover, from (\ref{cond1}), they have the
same diagonal elements.

5.\tab The non-diagonal elements of  $\rho^{12...N}_M$ are at
$(jd^i+1, kd^l+1)$ places with $jd^i<kd^l$. Now, $jd^i<kd^l$ may
happen in two ways: either $i<l$ or $j<k$ (when $i=l$). For $i<l$,
the non-diagonal element at $(jd^i+1, kd^l+1)$ is found to be
$w_{jd^i} \bar {w}_{kd^l}$ by comparing the coefficients of
$|0j\rangle\langle k0|$ from $\rho^{(N-l)(N-i)}_M$ and
$\rho^{(N-l)(N-i)}_W$. For $i=l$ ( and hence $j<k$), the same can
be achieved by comparing the coefficients of
$|0j\rangle\langle0k|$ from $\rho^{J(N-i)}_M$ and
$\rho^{J(N-i)}_W$ with any $J\ne i$. Thus $r_{(jd^i)(kd^l)}
=w_{jd^i}\bar{w}_{kd^l}$ and hence
$\rho^{12...N}_M~=~|W\rangle_N^d\langle W|$. \hfill$\Box$

\textbf{\emph{II. $N$-qubit generalized Dicke states}}:

The generalized Dicke states are defined by \bq \label{gdnl1}
\gdnl=\sum_ia_i|i\rangle \eq where $i=|i_1i_2...i_N\rangle$ and
the sum varies over all permutations of $\ell$ number of 1 and
$N-\ell$ number of 0. When all coefficients are equal, they are
known as Dicke states which have many interesting properties such
as permutational invariance, robustness against decoherence,
measurement and particle loss. Some important applications of
Dicke states include telecloning, quantum secret sharing,
open-destination teleportation and quantum games \cite{Dref2}. In
particular, implementation and various interesting applications of
$W$ states and their connection with Dicke states has been studied
in \cite{Wref1}. Thus Dicke states serve as a good test bed for
exploring multiparty correlations.

Let us first write the above state in (\ref{gdnl1}) as ( to have a
grip on the coefficients )\bq \label{gdnl2}
\gdnl=\sum\limits_{i_1>i_2>...>i_{\ell}=0}^{N-1}a_{2^{i_1}+2^{i_2}+...
+2^{i_{\ell}}}|B_N(2^{i_1}+2^{i_2}+...+2^{i_{\ell}})\rangle\eq
where $B_N(x)$ is the binary representation of the decimal number
$x$ in an $N$-bit string and $a_i$'s are arbitrary non-zero
complex numbers satisfying the normalization condition.

Retaining only the upper-half elements, we can write
\bq\label{gdnldens} \gdnl\langle GD_N^{(\ell)}|=\sum\limits_{i\le
j}g_{ij}|B_N(i)\rangle\langle B_N(j)|\eq where
$g_{ij}=a_i\bar{a}_j$ and $i,j$ vary over the decimal numbers
obtained by converting the $N$-bit binary numbers having $\ell$
number of 1 (and $N-\ell$ numbers of 0). In matrix form
$\gdnl\langle GD_N^{(\ell)}|$ will have non-zero entries
($g_{ij}$) only at ($i+1$, $j+1$) positions.

Since $1\le\ell\le N-1$, considering the entanglement, it is
sufficient to take $\ell\le\lfloor\frac{N}{2}\rfloor$=
\emph{Integral part} of $\frac{N}{2}$, as the states corresponding
to other $\ell$'s are LU-equivalent to these states. Any property
of these later states will follow from the corresponding former
states obtained by changing $0$ and $1$ throughout the bases. For
example, the two classes $|GD_N^{(N-2)}\rangle$ and
$|GD_N^{(2)}\rangle$ have the same property. We shall now prove an
interesting property of these states.

\textbf{Theorem 2 :}\emph{ For
$1\le\ell<\lfloor\frac{N}{2}\rfloor$, the generalized Dicke state
$\gdnl$ is uniquely determined by its $2\ell$-partite marginals.}

Note that we have excluded the case
$\ell=\lfloor\frac{N}{2}\rfloor$. The reason for this exclusion
will be described later. We will prove this theorem in two
parts---firstly we shall show that if any density matrix has the
same $(\ell+1)$-partite marginals as those of $\gdnl$, then it
must share the same diagonal elements with $\gdnl\langle
GD_N^{(\ell)}|$. But there will be some (off-diagonal) elements in
a general density matrix which will never appear in any
($\ell+1$)-partite marginal. So, to include these elements, we
have to consider RDMs of more parties. In the second part we will
show that it is sufficient to consider the $2\ell$-partite
marginals to prove the uniqueness (i.e., the two matrices share
the same off-diagonals).

\textbf{\emph{Proof}:}

1.\tab If possible, let there exist an $N$-qubit density matrix
(possibly mixed)\bq\label{rhom2}
\rho_M^{12...N}=\sum\limits_{i=0}^{2^N-1}\sum\limits_{j=i}^{2^N-1}r_{ij}|B_N(i)\rangle\langle
B_N(j)|\eq having the same $(\ell+1)$-partite marginals as those
of $\gdnl$. We shall prove that $\rho_M^{12...N}$ and
$\gdnl\langle GD_N^{(\ell)}|$ share the same diagonals.

2.\tab We will first consider the diagonal elements of RDMs. Since
there is exactly $\ell$-number of non-zero entry (each is 1) in
every basis term of $\gdnl$, the coefficient of
$$|i_1i_2...i_{\ell+1}\rangle\langle i_1i_2...i_{\ell+1}|$$ in any
$(\ell+1)$-partite marginal should be zero in which every $i_k$ is
1. This constraints the form of $\rho_M^{12...N}$ in eqn
(\ref{rhom2}) to have some coefficients ($r_{ij}$) vanishing. In
(\ref{rhom2}), only those $r_{ij}$ will be non-zero for which both
of $B_N(i)$ and $B_N(j)$ has at most $\ell$ number of 1. We shall
now show that only those $r_{ii}$ in (\ref{rhom2}) are non-zero
for which $B_N(i)$ has exactly $\ell$-number of 1.

a).\tab Let us consider the coefficient of
$|i_1i_2...i_{\ell+1}\rangle\langle i_1i_2...i_{\ell+1}|$ where
$\ell$ number of $i_j$'s are 1 and only one is zero, in the RDM of
some parties $J_1,~J_2,...,J_{\ell+1}$. There is exactly one basis
term in $\gdnl$ having $i_k$ at $J_k$th place, with the
coefficient $a_i$ where $i=2^{N-J_{\ell+1}}+...+2^{N-J_1}$. So,
when the RDM is calculated from $\gdnl$, the coefficient is
$g_{ii}=|a_i|^2$. Again, there is exactly one non-zero $r_{ii}$ in
(\ref{rhom2}) such that $B_N(i)$ has $i_k$ at $J_k$th place (since
$B_N(i)$ can have at most $\ell$ number of 1, in order for
$r_{ij}\ne0$). Therefore, comparing the coefficients (of this
term), $r_{ii}=g_{ii}$. Considering all permutations of this term
and all possible set of $(\ell+1)$-number of parties, it follows
that $r_{ii}=d_{ii}$, for all decimal $i$ so that $B_N(i)$ has
$\ell$-number of 1.

b).\tab Now we will show that all other $r_{ii}$, corresponding to
which $B_N(i)$ has less than $\ell$  number of 1, should be 0.
First consider those $r_{ii}$'s corresponding to which $B_N(i)$
has $(\ell-1)$ number of 1. Then comparing the coefficients of
$|i_1i_2...i_{\ell+1}\rangle\langle i_1i_2...i_{\ell+1}|$ where
$(\ell-1)$ number of $i_j$'s are 1, from the RDMs (considering all
possible set of parties and using the result of step a) above), we
get $r_{ii}=0$. Similarly, all other $r_{ii}$'s, corresponding to
which $|B_N(i)\rangle$ has less than $\ell$ number of 1, should be
zero. Finally from normalization ($\sum r_{ii}=\sum g_{ii}=1)$, it
follows that $r_{00}=0$.

Thus, collecting the results of a) and b) it follows that
$\rho_M^{12...N}$ in (\ref{rhom2}) reduces to the same form as
$\gdnl\langle GD_N^{(\ell)}|$ in (\ref{gdnldens}) and they have
the same diagonal elements $r_{ii}=g_{ii}$. The only remaining
task to prove the uniqueness is to show that they have the same
non-diagonal elements too.

3.\tab Consider a non-diagonal element $r_{ij}$ with
$i=|...i_1..i_{\ell}...\rangle$ and
$j=|...j_1..j_{\ell}...\rangle$. (each of $i_k$ and $j_k$ is 1).
Since $\ell<\lfloor\frac{N}{2}\rfloor$, there will be some terms
$|i\rangle\langle j|$ in the density matrix, the coefficients
($r_{ij}$ or $a_i\bar{a}_j$) of which will never occur in any
$(\ell+1)$-partite marginal. For example, the coefficient of
$|000\ldots011\rangle\langle110\ldots0|$, or
$|010\ldots01\rangle\langle100\ldots010|$ will never appear in any
tripartite marginal. Generically, those $r_{ij}$'s $(i<j)$ for
which the Hamming Distance between $B_N(i)$ and $B_N(j)$ is
greater than $\ell+1$, will never occur in any $(\ell+1)$-partite
marginal; because partial tracing over the remaining parties will
yield 0. Thus, the elements $r_{ij}$'s with
$j=\bar{i}$=\emph{complement} of $i\equiv2^N-1-i$ (these are the
elements on the secondary diagonal of the density matrix) will
never occur.

Since these $r_{ij}$'s never occur in any $(\ell+1)$-partite
marginal, these are unconstrained elements (i.e., these can take
any values and need not be $a_i\bar{a}_j$, which is required for
the uniqueness of the two density matrices). So, there exists an
infinite number of $2^N\times2^N$ hermitian, unit-trace matrices
sharing the same diagonals and $(\ell+1)$-partite marginals with
$\gdnl\langle GD_N^{(\ell)}|$. However all such matrices may not
be valid density matrices because of the semi-positivity
restriction ($\rho\ge0$). So, for some particular choices of the
coefficients $a_i$'s, there may (or may not) exist a valid density
matrix other than $\gdnl\langle GD_N^{(\ell)}|$. Therefore, there
is an ambiguity about the general case: what is the minimum number
of parties whose RDM's can generically determine the $\gdnl$ state
uniquely?

4.\tab To answer this question, we observe that the possible
maximum Hamming Distance between $B_N(i)$ and $B_N(j)$ is $2\ell$.
Therefore, if we consider the $2\ell$-partite marginals, each
$r_{ij}$ will appear in some RDM's and hence should be constrained
to satisfy some relation with $a_i$'s. We shall now show that
considering $2\ell$-partite marginals, indeed yield
$r_{ij}=a_i\bar{a}_j$.

To prove it, consider a non-diagonal element $r_{ij}$ with
$i=|i_1i_2\ldots i_N\rangle$ and $j=|j_1j_2\ldots j_N\rangle$. Let
the $\ell$ 1's in $i$ be at $I_k$th places (counting from left to
right) and those in $j$ are at $J_k$th places. If the two sets
$\{I_k\}$ and $\{J_k\}$ are different (i.e.,
$\{I_k\}\cap\{J_k\}=\Phi$), then we get a set of $2\ell$ number of
parties $\{I_k,J_k\}$ and we can arrange all $I_k$ and $J_k$'s (since
$I_k, J_k \in \{1,2\ldots,N\}$) in increasing order. Let us called
them $\{P_k\}$ (i.e., $P_1<P_2<\ldots<P_{2\ell}$). If
$\{I_k\}\cap\{J_k\}\ne\Phi$, we can add any number(s) from $\{1,2,\ldots,N\}$
to the set $\{P_k\}$ (maintaining the order) so that it contains $2\ell$ number
of elements. Let $s_k$ be the $P_k$th bit (from left to right) in $B_N(i)$ and
those in $B_N(j)$ be $t_k$. Then comparing the coefficient of $|s_1s_2\ldots
s_{2\ell}\rangle\langle t_1t_2\ldots t_{2\ell}|$ from the RDMs
$\rho^{P_1P_2\ldots P_{2\ell}}$, it follows that $r_{ij}=a_i\bar{a}_j$
and hence the proof.
\hfill$\Box$

\textbf{\emph{Remark 2}:} It is worth mentioning that the Theorem
2 can be viewed as a sufficient condition. It states that it is
sufficient to consider the $2\ell$-partite marginals to determine
$\gdnl$. However, it may happen (e.g., for some specific state in
this class) that the state $\gdnl$ can be determined from fewer
than $2\ell$-partite marginals. In this sense, we do not know
whether this is an optimal bound. We have used the $2\ell$-partite
marginals to drive out the possibility of presence of another
density matrix having different off-diagonals but sharing the same
diagonals. The off-diagonals $r_{ij}$ are arbitrary but are
constrained to satisfy the requirement that the resulting matrix
should be PSD. This automatically puts some restrictions on the
off-diagonals e.g., $|r_{ij}|\le\sqrt{r_{ii}r_{jj}}$. There is a
possibility of reducing the number of parties using some further
properties of density (PSD) matrices (or using some different
techniques). In the present technique, $2\ell$ partite marginals
are sufficient. A limitation of the present technique is that if
the maximum Hamming distance (between the bases) is $N$, then it
gives the trivial result. In the case
$\ell=\lfloor\frac{N}{2}\rfloor$, for odd $N$, the technique
produces the result of \cite{WL} and for even $N$, it gives no
useful result. Thats why we have excluded this case in Theorem 2.

Another interesting issue is the number of RDMs needed to identify
a state. For example, it has been shown by Diosi \cite{others1}
that among pure states, only two (out of three) bipartite
marginals are sufficient to determine a generic three qubit pure
state ($GHZ$ and its LU equivalents are the only exception). If we
restrict ourselves only to pure states , then the number of RDMs
can be considerably reduced. The result is stated more precisely
in the following theorem.

\textbf{Theorem 3 :}\emph{ Among arbitrary pure states, the
generalized Dicke state $\gdnl$ is uniquely determined by its
$(\ell+1)$-partite marginals. Moreover, only $^{N-1}C_{\ell}$ (out
of $^NC_{\ell+1}$) number of them having one party common to all,
are sufficient.}

\textbf{\emph{Proof}:}

Let us take the first party as the common one and consider the following
RDMs $\rho^{1i_2i_3\ldots i_{\ell+1}}$. The proof for diagonal part
has already been proved in the first part of Theorem 2. The proof for
the non-diagonal part follows by comparing the coefficients of $|B_{\ell+1}(i)\rangle
\langle B_{\ell+1}(j)|$, where $|B_{\ell+1}(i)\rangle$ and $|B_{\ell+1}(j)\rangle$
has exactly $\ell$ number of 1s. \hfill$\Box$

\textbf{\emph{Remark 3}:} We wish to mention here that as we are
considering the most general class of Dicke states, it is not
possible to determine the states from fewer than
$(\ell+1)$-partite marginals. It may happen that for some specific
choices of the coefficients, $\gdnl$ is uniquely determined from
fewer than $(\ell+1)$-partite marginals, but in general, not all
states can be determined. For example, the following two states
\bqa ~~~~~|GD_4^{(2)}\rangle &=&
r_3e^{i\theta_3}(|3\rangle+|12\rangle)
+r_5e^{i\theta_5}(|5\rangle+|10\rangle) +
r_6e^{i\theta_6}(|6\rangle+|9\rangle)\nonumber\\ \mbox{and~}
|GD_4^{(2)'}\rangle &=& r_3e^{-i\theta_3}(|3\rangle+|12\rangle)
+r_5e^{-i\theta_5}(|5\rangle+|10\rangle) +
r_6e^{-i\theta_6}(|6\rangle+|9\rangle) \nonumber\eqa are not
determinable since they share the same bipartite marginals.
[$r_i,~\theta_i$ are real and the base $|x\rangle$ should be read
as $|B_4(x)\rangle$].

\textbf{\emph{III. Generalization to $d$-dimension}}: The
generalized $d$-dimensional Dicke states are defined by
\bq\label{ggd1} |D_N(k_0,
k_1,...,k_{d-1})\rangle=\sum_ic_i|i\rangle\eq where \bq
\label{cggd1}
i=|\underbrace{0...0}_{k_0}\underbrace{1...1}_{k_1}....
\underbrace{\overline{d-1}...\overline{d-1}}_{k_{d-1}}\rangle\eq
and the index $i$ varies over all possible different permutations
of $k_0$ number of 0, $k_1$ number of 1,..., $k_{d-1}$ number of
$\overline{d-1}$; $k_0+k_1+...+k_{d-1}=N$. These states are
genuinely entangled. Using the same technique as in the proof of
Theorem 2, we can prove the following result about the reducible
correlations in these states.

\textbf{Theorem 4 :}\emph{ If $K (\equiv2\ell<N)$ be the maximum Hamming distance between
the bases (\ref{cggd1}), the state given by (\ref{ggd1}) is uniquely determined by its
$K$-partite RDMs.}

As an example, any state of the class $|D_{2009}(2004,2,3)\rangle$
is determined by its $10$-partite RDMs.

\textbf{\emph{IV. Quantum Marginal Problem}}: The basic issue
concerning the Quantum Marginal Problem (QMP) is the following:
does there exists a joint quantum state consistent with a given
set of RDMs? It is known that a general solution to the QMP would
provide a solution to the \emph{N-representability problem} in
quantum chemistry, e.g., to calculate the binding energies of
complex molecules \cite{bookqmp}.  A particular class of  QMP is
`Symmetric Extension', which has direct application in Quantum Key
Sharing, Quantum Cryptography etc \cite{Symextn}. Although plenty
of literature is available \cite{paperqmp}, there is no general
method to get the exact solutions. One needs to calculate the
marginals from an arbitrary state (which is the expected joint
state) and then compare with the given ones. For large number of
marginals, the problem becomes very difficult as we need to solve
several complex equations.  However, if there is some symmetry in
the RDMs (e.g.,  for $W$-states, Dicke states , all RDMs have
similar form with many vanishing elements), the technique
presented in our work can be applied to find a solution. As an
interesting example, it was mentioned in \cite{LPW} that for the
set of RDMs $\{\rho^{AB},~\rho^{BC},~\rho^{AC}\}$ where (in
computational
basis)\bq\rho^{AB}=\rho^{BC}=\rho^{AC}=|\psi^-\rangle\langle\psi^-|
\eq (with $|\psi^-\rangle = (|01\rangle - |10\rangle)/\sqrt{2}$)
there exists no consistent 3-qubit state. We can prove this easily
using our technique. If possible, let \bq
\rho^{ABC}~=~\sum_{i=0}^{7}\sum_{j=i}^{7}r_{ij}|B_3(i)\rangle\langle
B_3(j)|\eq where $B_3(x)=$`Binary representation of $x$ in a 3-bit
string', has the given marginals. Then comparing the first and
last diagonal elements (i.e., the coefficients of
$|00\rangle\langle00|$ and $|11\rangle\langle11|$) of the RDMs, we
get $r_{ii}=0~\fa i=0(1)7$, an impossibility!

\textbf{\emph{Conclusions}}: Though the general framework is still
far way, through this work we have made considerable progress
towards understanding the nature of reducible correlations. It has
been shown that the correlations in some classes of multipartite
states can be reduced to lower order ones. This provides some
insight into the characterization of multiparty entanglement, such
as the determination of generalized $W$ state from its bipartite
RDMs proves that the entanglement therein is necessarily of
bipartite nature. The large class of generalized Dicke states
$\gdnl$ have shown to be determined by their $2\ell$-partite
marginals, where $1\le \ell <\lfloor \frac{N}{2}\rfloor$. Thus,
these states have information at the most at $2\ell$-partite level
and it can not be reduced beyond $(\ell+1)$-partite level.

In general, the entangled states which are determined by their
$K$-party RDMs, can be used as resources for performing
information related tasks, specially if some of the parties do not
cooperate. In such situations, it is not necessary that each party
cooperates with all others; cooperation with only $K-1$ parties is
sufficient. The $K$-partite residual entanglement would serve the
purpose. For example, because of the bipartite nature of
entanglement, the $N$-qubit $W$-state is very robust against the
loss of $(N-2)$ parties.

Recently it has been shown that $(N-1)$-qudit RDMs uniquely
determine the Groverian  measure of entanglement of the $N$-qudit
pure state \cite{EJ}. So it is likely that for the pure states,
which are determined by their $K$-partite RDMs, the entanglement
measure may be characterized by these RDMs. However, this requires
further investigation.

Finally, we have shown by an example that our approach can be
applied to Quantum Marginal Problem, at least for simple
(low-dimensional) cases.

\end{document}